\documentstyle[epsf,12pt]{article}
\newcommand{\be}{\begin{equation}}
\newcommand{\ba}{\begin{eqnarray}}
\newcommand{\ee}{\end{equation}}
\newcommand{\ea}{\end{eqnarray}}
\newcommand{\nn}{\nonumber}

\newcommand{\eV}{\;{\mathrm eV}}

\newcommand{\GeV}{\;{\mathrm GeV}}

\newcommand{\cm}{\;{\mathrm cm}}
\newcommand{\sterad}{\;{\mathrm sterad}}
\newcommand{\ergs}{\;{\mathrm ergs}}
\newcommand{\simlt}{\stackrel{<}{{}_\sim}}
\newcommand{\simgt}{\stackrel{>}{{}_\sim}}
\newcommand{\ol}{\overline}

\newcounter{currequation}
\begin{document}
\title{On the possible dark-matter content of QSO's and of compact, very
massive entities in the nuclei of galaxies: a metastable particle with
mass of about $10^{10}\GeV$}
\author{Saul Barshay and Georg Kreyerhoff\\
III.~Physikalisches Institut A\\
RWTH Aachen\\
D-52056 Aachen, Germany}
\maketitle
\vspace{-9cm}
\flushright {PITHA 98/31}
\vspace{8cm}\par
\abstract{We examine some general astrophysical results which can be related
to the hypothesis that very heavy, metastable particles constitute compact, very
massive central entities in QSO's and the core of galaxies. The mass and lifetime
have been calculated in detail previously: the mass is about $10^{10}\GeV$; the lifetime
is $\simgt 10^{21}\;\sec$. The specific decay gives rise to a new source of very
large amounts of energy in radiation. The essence of the ideas discussed in this
paper is that very massive, metastable dark matter constitutes entities near to
black-hole conditions, and that it is decay which provides a large primary energy source
from such entities, as components of QSO's, AGN, and possibly GRB's.}
\section{Introduction}
The purpose of this paper is to consider the hypothesis that very heavy, metastable
particles\cite{ref1}, which can make up the main part of cold dark matter\cite{ref1},
are present in, and give rise to the principle sources of energy in compact, very
massive astrophysical entities. The latter are observed as quasi-stellar objects (QSO's) and as
possibly related entities in the core of galaxies. Such entities can be considered
to contain the mass equivalent of about $10^{10}$ solar masses\footnote{In this
paper, we use the units of high-energy particle physics. A solar mass made up
of about $10^{57}$ nucleons is about $10^{57}\GeV$.} within a domain characterized
by a linear dimension of the order $10^{16}\cm$ ($\sim$ a light week)\cite{ref2}.
We have a specific motivation for examining some interrelated results which can follow
naturally from the hypothesis. The motivation is provided by the results of detailed
calculations\cite{ref1,ref3} using the renormalization group equations in a
(chiral-like) dynamical, cosmological model, involving only a neutral scalar inflaton
field\footnote{We have calculated explicit potentials which exhibit features characteristic
of an inflationary period in the early universe\cite{ref1,ref3}. These potentials have 
both a calculated maximum and a calculated minimum. The scalar field ``rolls'' from
the maximum to the minimum. The inflaton is the metastable, massive quantum of the
scalar field; the number density is related to field fluctuations about the minimum\cite{ref1}.}, 
a neutral pseudoscalar
field, and a neutral lepton. Two new ideas emerge from the results.
\begin{itemize}
\item[(1)] A large part of the dark matter in the present universe can be composed
of inflatons, whose mass is calculated to be near to $10^{10}\GeV$. The oft-repeated
statement that the inflaton mass must be $\simgt 10^{13}\GeV$ is based upon the assumption that rapid 
inflaton decay produces the radiation composed of ordinary matter. There is no empirical
basis for this assumption. In our model \cite{ref1}, the inflaton is decoupled from ordinary
matter (but for gravitational effects). The radiation is produced by decay processes
involving other primary particles produced by the
release of vacuum energy. The small CMB anisotropy can be related to the very small
parameter in the inflaton potential, $\sqrt{\lambda}\sim 10^{-7}$. (The smallness of $\lambda$
can have a natural dynamical origin \cite{ref1}.)
\item[(2)] The inflatons are not absolutely stable; they can decay in a specific
way, with a lifetime calculated to be several orders of magnitude greater than the present
age of the universe, $t_0\sim 4\times 10^{17}\sec$. 
\end{itemize}
In the next sections, we consider some qualitative consequences of the ideas taken together
with the above hypothesis.
\section{Degrees of freedom}
In the astrophysical context, we consider as degrees of freedom the large numbers
which are typical for the number of nucleons in the universe, the number of nucleons
in a star$^{F1}$, the number of stars in a galaxy, the number of galaxies, and
the large dimensions which characterize certain relevant scales of structure:
$\sim 10^{16}\cm$ for compact, massive central entities
(to $\sim 10^6$ for neutron stars), $\sim 10^{23}\cm$ for galaxies, $\sim\sqrt{2}\times 10^{28}\cm$
for the universe accessible to observation\footnote{For $h_0^2\cong \frac{1}{2}$}. Further,
we consider the energy density in cold dark matter, and that in nucleons, as fractions of the
critical density $\rho_c\cong 4\times 10^{-47}\GeV^4\cong 0.5\times 10^{-5}\GeV/\cm^3$.$^{F3}$
There are interesting possible relationships between these quantities which follow
from the hypothesis, and the quantitative results (1) and (2), stated in the introduction.\par
In the numerical estimates in this paper, we use for the mass of the inflaton one of the
specific values calculated in our recent work\cite{ref1}, $m\cong 5\times 10^{10}\GeV$.\footnote
{Another specific value calculated in reference 1 is $m\cong 6\times 10^9\GeV$. We comment
below on the effects of a somewhat larger value than is explicitly used here.}  Assume
that one compact, massive object composed of about $10^{57}$ inflatons exists
in the core of a typical galaxy.\footnote{Concerning the implicit assumption that about
each galaxy is associated with a compact, massive central entity, there are at least
two reasons for fewer to be observed by the techniques mainly employed. One is the
usual argument that the sources were once active, but have ``expired'' (due to eventual
lack of ``fuel'' i.~e.~infalling charged matter) over a relatively short time span
\cite{ref2}. In the context of the ideas in this paper, a second reason could be that the
sources never ``lit up'' because of lack of ordinary charged matter associated with the
central, dark-matter structure. However, they would still be emitting very high-energy
neutrinos from inflaton decay, as discussed in section 3. An accretion region of decay neutrinos
is possible; conversion to electrons would imply radiation from ejected, as well as infalling, charged matter.} The central
object is thus a compact entity with the equivalent of about $10^{10}$ solar masses. This is
a mass equivalent near to that of some QSO and galactic-core entities. 
Assume that the galactic mass of all of the stars plus the mass of gaseous 
matter not in stars, a total mass
essentially constituted from nucleons of mass $\sim 1\GeV$, is approximately the same
as the mass of the central entity. The number of nucleons typically present
in a star is $\sim 10^{57}$. The number of nucleons in gaseous material is of the
order of five times that in all stars \cite{ref4}. Denote a typical number of stars
in a galaxy by $s$. Then $s$ is determined from
\be
(5\times 10^{10}\GeV)\times (10^{57}) \cong (1\GeV)\times (5s)\times(10^{57})\to s\cong 10^{10}
\ee
Denote a typical number for all galaxies by $g$. The total number of nucleons is about
\be
n_b \cong 10^{-10}\times (10^{88})\cong 10^{78}
\ee
where $10^{88}$ is approximately the total number of photons, and $\sim 10^{-10}$ is the
small empirical number which characterizes the baryon asymmetry. The galaxy number is determined
from 
\be
10^{78}\cong g (5s) \times (10^{57}) \cong g (5\times 10^{67})\to g\cong 2\times 10^{10},\;\; g\times s \cong
2\times 10^{20}
\ee
(We are not considering a possible large number of ill-formed, irregular dark entities.)
The energy density of baryonic matter is
\be
\rho_b \cong \frac{10^{78}(1\GeV)}{\frac{4\pi}{3}\left(\sqrt{2}\times 10^{28}\cm\right)^3}\cong
0.85\times 10^{-7}\GeV/\cm^3\cong 1.7\%\times \rho_c
\ee
This is essentially the same number as that produced by the recent, empirically informed, universal
baryon budget-keeping\cite{ref4}.\par
We can estimate an energy density in dark-matter inflatons. Paralleling the situation for
nucleons, assume that the number of inflatons in less compact configurations 
is about ten times that in the core-entity, i.~e.~$\sim 10\times 10^{57}$. With $g=2\times 10^{10}$
from eq.~(3)$^{F5}$, the cold dark-matter energy density from inflatons is
\ba
\rho_{\mathrm{CDM}}&\cong&\frac{(5\times 10^{10}\GeV)\times(10^{58})\times (2\times 10^{10})}
{\frac{4\pi}{3}\left(\sqrt{2}\times 10^{28}\cm\right)^3}\cong 0.85\times 10^{-6}\GeV/\cm^3\nn\\
&\cong&17\% \times \rho_c
\ea
This is about ten times the energy density of baryons; it remains however, a small fraction of the
critical density\footnote{We do not believe in the, often repeated, popular mythology that inflation
in the early universe necessarily implies that a critical density must be observed today. In particular,
insistence upon trying to make results practically independent of inital conditions, appears to
be a prejudice. (It does not acquire credibility in an empirical science through use of jargon,
like ``generic''.)}. We
also estimate the dark-matter energy density on the scale of a single galaxy, using a galactic
dimension of $\sim 10^{23}\cm$.
\be
\rho_{\mathrm{CDM}}^{\mathrm{galaxy}}\cong\frac{(5\times 10^{10}\GeV)\times (10^{58})}
{\frac{4\pi}{3}(10^{23}\cm)^3}\cong 0.125\GeV/\cm^3
\ee
This is sufficient to meet the rotation-curve anomaly. Note that increasing $m$ by about
3 increases a typical star-number per galaxy to $s\sim 3\times 10^{10}$ 
and increases $\rho_{\mathrm{CDM}}^{\mathrm{galaxy}}$
to $\sim 0.38\GeV/\cm^3$. Thus, $m$ should lie within a relatively small interval around
$10^{10}\GeV$\cite{ref1}.$^{F4}$\par
Years ago, in his textbook\cite{ref5}, Hoyle noted a numerical relationship, with the
remark: ``Here is something odd to think about''. Hoyle observed that the ratio of the
total number of nucleons to the dimension characteristic of the universe is nearly the same
large number as the ratio of the number of neutrons to the dimension characteristic of
a compact neutron star, $\sim 10^6\cm$. That is, we have
\be
\frac{10^{78}}{(\sqrt{2}\times 10^{28})\cm}=0.7\times 10^{50}\cm^{-1}\sim\frac{10^{57}}{10^6\cm}=10^{51}\cm^{-1}
\ee
Although neutron stars could exist with appreciably less than $10^{57}$ neutrons, this near
equality ``specifies the magic $10^{57}$''\cite{ref5}. Since the mass of the nucleon
is $\sim 1\GeV$, this relationship is equivalent to a kind of ``column energy''
(denote this by $\epsilon$) characterized by a quantity of the order of $\epsilon_b\sim 10^{51}\GeV/\cm
\cong (2\times 10^{37})\GeV^2$. This quantity is almost the square of the Planck mass,
$M_P^2\cong 1.5\times 10^{38}\GeV^2$. The near-equality can be viewed as relating
a particle-physics scale $(\sim 1\GeV)$ to a cosmological scale.
Now consider such a quantity for the inflaton dark matter. On the dimension scale of
the universe, with $\sim (10\times 10^{57})=10^{58}$ inflatons in the core entity and
gas of approximately each of $\sim 2\times 10^{10}$ galaxies, we have for the inflaton
\be
\epsilon_i\cong\frac{(5\times 10^{10}\GeV)\times (10^{58})\times (2\times 10^{10})}{(\sqrt{2}\times 10^{28})
\cm} = 7\times 10^{50}\GeV/\cm
\ee
Consider that a single compact, massive dark-matter central object is characterized by a dimension
of the order of light-week, $\sim 2\times 10^{16}\cm$. Then, on the scale of this
dimension for a QSO or for a galactic-core entity, we have
\be
\epsilon_i^{\mathrm{core}}\cong\frac{(5\times 10^{10}\GeV)\times (10^{57})}{(2\times 10^{16})\cm}
= 2.5\times 10^{51}\GeV/\cm
\ee
So $\epsilon_i^{\mathrm{core}}\cong \epsilon_i\cong M_P^2$, or turning the argument around,
requiring the near-equality specifies the dimension of about a light-week. This is just
above the Schwarzschild radius.
With consideration given to the numbers resulting from eqs.~(1,3,5,6,8,9), we conclude
this section by remarking that the repeated presence of the number $\sim 10^{10}$ among
basic astrophysical quantities - the approximate star number per galaxy, the approximate
galaxy number, and the possible (maximal) solar mass equivalent in QSO's and galactic-core 
entities - points to the possible effects of a metastable particle with mass near to
$10^{10}\GeV$, i.~e.~about $10^{10}$ times the nucleon mass. Aggregates of this particle
have primarily  gravitational interactions.\footnote{There is a repulsion between inflaton
quanta originating in the very small self-coupling \cite{ref1}. The internal energy might be
sufficient to balance gravity only beyond galactic dimension.}
\section{Energy considerations}
The first observation concerning energy is simply the correct order of magnitude of the
very large gravitational potential energy associated with the compact, massive dark-matter core. 
The (classical) order
of magnitude of the self-energy is
\be
-V_{\mathrm{gravity}}^{\mathrm{inflaton}}\sim \frac{3}{5}\frac{(5\times 10^{10}\GeV \times 10^{57})^2}
{(10^{19}\GeV)^2(10^{16}\cm)}\cong 3\times 10^{67}\GeV\cong 4.8\times 10^{64}\ergs
\ee
This is somewhat greater than an empirical energy equivalent in emission from a very
strong radio source. Such entities would have a tendency to form at an earlier stage than usual
structure formation. (This could be about $10^6\sec$.) They could give a kind of impulse
for later galaxy formation\cite{ref6a}.\par
Here we are concerned with a completely new source of a comparably large quantity of energy
in radiation, which results from the specific decay of the inflaton, with a very long lifetime.
The inflaton decays only into a neutrino-antineutrino pair, with a lifetime which we have estimated
to be of the order of $10^{21}\sec\sim (2.5\times 10^3)t_0$, or greater\cite{ref1}. (This occurs because
of the very small mixing of the massive neutral lepton, to which the inflaton couples, with
the heaviest of the known neutrinos.) The decay neutrino and antineutrino have energy of the order
of $10^{10}\GeV$. In a gas of these neutrinos, interactions occur which give rise
to lepton-antilepton pairs and quark-antiquark pairs. In addition,
if there is an existing gas of electrons and nucleons in or nearby the system, then the pressure
of the very high-energy neutrinos can accelerate these electrons. Prior to decays,
the total energy in inflaton mass associated with the compact, central entity is
$\sim (5\times 10^{10}\GeV)\times (10^{57})\sim 8\times 10^{64}\ergs$.
If one considers less dense configurations, of about $(10\times 10^{57})$ inflatons
(as in eqs.~(5,6)) from a dimension of the order of $10^{17}\cm$ (beyond the central entity
with dimension of $\sim 10^{16}\cm$), then the number density is
\be
n_i\sim\frac{10^{58}}{\frac{4\pi}{3}(10^{17}\cm)^3}\cong 2.5\times 10^6/\cm^3
\ee
This number is not unusual; it is a particle density characteristic of ordinary matter
in some gaseous layer away from the center of a QSO. A small fraction of the
inflatons have decayed, of the order of $(t_0/\tau)\cong 0.4\times 10^{-3}$. (In the following
numerical estimates we use a little smaller fraction, $\sim \frac{1}{8}\times 10^{-3}$.
\footnote{The numerator is generally a time less than $t_0$. This reflects the subtraction
from $t_0$ of the ``look-back'' time to the entity.}) The energy in neutrinos is then
about $10^{62}\ergs$, with a pair number-density of $\sim 3\times 10^2/\cm^3$.
These very high-energy neutrino-antineutrino pairs will interact and produce very
energetic charged leptons and antileptons, via $\nu+\ol{\nu}\to \ell^-, \ell^+$.
(Also, quark-antiquark pairs.) If we assume relevant cross sections of roughly $\sigma_\nu
\sim 10^{-32}\cm^2$, then a mean-free path of $\sim\frac{1}{3}\times 10^{30}\cm$
results in a fraction $\sim 1.5\times 10^{-13}$ of the  number of neutrino pairs
converting to $\ell\ol{\ell}$ (and $W\ol{W}$, $Z\ol{Z}$) over an interaction-path length
of $\sim 10^{17}\cm$. The energy in promptly-produced\footnote{Electrons
are also produced in (time dilated) decays following $\tau^-\tau^+$ production.} 
$e^-,e^+$ (via $Z$'s, $W$'s),
is then $(\sim 10^{-13})(10^{62}\ergs)\cong 10^{49}\ergs$.
Distributing this over a galactic dimension gives an energy density contained
in the very high-energy $e^-$ and $e^+$ of the order of
\be
\rho_{e^+e^-}\sim \frac{10^{49}\ergs}{\frac{4\pi}{3}(10^{23}\cm)^3}\cong
(0.25\times 10^{-6})(10^{-14}\ergs/\cm^3)
\ee
This estimate is interesting for the following reason. The number $10^{-14}{\mathrm{ergs}}/\cm^3$
is similar to a (total) energy density characteristic of our galactic cosmic-ray particles
($\sim 10^{-12}\ergs/\cm^3$), times the approximate $1\%$ of this in electrons. The number of
order $10^{-6}$ is roughly indicative of the fraction in electrons with energies a few
times $10^{10}\GeV$. This number is at most the small fraction $\sim(10^6/10^{11})=10^{-5}$
which arises from the empirical\cite{ref6}, approximate effective $(1/E^2)$ fall-off
of the cosmic-ray particle flux between $\sim 10^6\GeV$ and $\sim 10^{11}\GeV$.\par
Similar numbers to those above, hold for the transfer of energy from very high-energy
neutrinos to electrons via scattering (and to nucleons, via deep-inelastic scattering),
if there is a nearby gas of electrons and nucleons with number density at least as large
as that of the neutrinos from inflaton decay. The important physical difference is of
course that, whereas neutrino annihilation produces positrons as well as electrons,
and quark-antiquark pairs, the neutrino ``pressure'' accelerates ``fuel'' (and does
not transfer all of the neutrino energy).
If some of these particles can ``escape'' the influence of the inner core, then very energetic
electrons and nucleons, as well as the neutrinos themselves from inflaton decay\cite{ref7},
can occur throughout the galaxy. (There would also be $\gamma$-rays from quark fragmentation.) 
\par
For the purpose of orientation, we give estimates for the flux of these very high-energy
neutrinos, and electrons, upon the earth's atmosphere, assuming no large losses
within the galaxy. An approximate flux formula gives\footnote{For $\tau \gg t_0$, and
$E_{\nu_\tau}\simlt {\frac{m}{2}}$. A similar estimate is made in ref.~8, using
possible distant, diffuse sources. Then $L\cong 10^{28}\cm$; this is nearly compensated
by a decrease in the diffuse $\rho_{\mathrm CDM}$ by about $10^{-5}$ \cite{ref1}. There
is a small dimunition in neutrino energy here, due to red-shift $z<1$. Diminished inflaton
decay at very early times tends to reduce contributions (at lower energies here) from
very large $z$.}
\be
I_{\nu,\ol{\nu}}\cong\frac{\rho_{\mathrm{CDM}}^{\mathrm{galaxy}}}{4\pi m}\frac{L}{\tau}
\cong 2\times 10^{-13}\;(\cm^2-\sec-{\mathrm{sterad}})^{-1}
\ee
We have used $\rho_{\mathrm{CDM}}^{\mathrm{galaxy}}\cong 0.125\GeV/\cm^3$ from eq.~(6),
$L\cong 10^{23}\cm$, and for $m \sim 5\times 10^{10}\GeV$ \cite{ref1},
$\tau\cong 10^{23}\;\sec$. Neutrinos ($\nu_\tau$) with
energies of about $10^{10}\GeV$ can have a cross section in air greater than
$10^{-33}\cm^2$. With an interaction probability in the atmosphere\cite{ref7}
of about $10^{-7}$, the effective flux becomes of the order of $10^{-20}
\;(\cm^2-\sec-{\mathrm{sterad}})^{-1}$.\footnote{
If the mass of $\nu_\tau$ is only $\sim 0.05\eV$ (instead of $\sim 1.8\eV$ as used in \cite{ref1}),
then the lifetime is lengthened by $\sim 10^3$, and the flux in eq.~(13) is reduced to $\sim 2\times 10^{-16}
(\cm-\sec-\sterad)^{-1}$. However, some compensation could occur from a possible
neutrino interaction probability in air as large as $\sim 10^{-5}$.} 
It is worth noting that cosmic-ray experiments have recently
reached exposures\cite{ref8} of $2.6\times 10^{20} \;(\cm^2-\sec-{\mathrm{sterad}})$, and
will eventually go much further. There are a few unusual events near to $10^{11}\GeV$\cite{ref7,ref8}.\par
Quasars may have a maximum occurrence near to red-shift $z\sim 2$. Their decline at later
times may be related to a dimunition of atmospheric ``fuel'' \cite{ref2}. However,
their decline at very early times may be related to the decreasing amount of energetic
neutrinos (and of conversion electrons$^{\mathrm{F5}}$) from inflaton decay.
\section{Summary}
In one of the concluding paragraphs of his book\cite{ref9}, Hoyle made the comment:
``There could be a connection here with the outbursts of radio galaxies and QSO's\ldots
I have had for some years the lurking suspicion that the cascades of highly energetic
particles responsible for our observations might be generated by the decays of some
superparticle.'' The considerations in this paper suggest that this may be part of
the truth. The other more traditional part, the very large energy source from gravitational
attraction is here also closely linked to the hypothetical massive particle.\par
Definite observational tests of these ideas are possible. In particular, through determination
of the amount of cold dark matter that is actually present\footnote{Our estimates for the inflaton
mass suggest a restricted range, from $\sim 10\times \rho_b$ to $\sim\frac{1}{2}\times\rho_c$.\cite{ref1}}
, and through the detection of very high-energy neutrinos. There is accumulating evidence
for the presence of massive dark objects in the centers of galaxies\cite{ref10}. This is
also relevant for the very high-energy cosmic rays, because if neutrinos from decay
of massive dark matter are emanating from these objects, then scattering in a dilute
atmosphere can produce energetic protons and gamma-rays. These could constitute a very
energetic component of hadron-like cosmic rays, from sources in the sky which are not very distant.\par
It is possible that a gamma-ray burst is powered by a very compact source
($\sim 10^8\cm$) of the very high-energy neutrinos from inflaton decays.
An energy of about $10^{52}$ ergs in relativistic motion would be produced by a source containing the equivalent of about $10^2$ solar masses (only about
$10^{49}$ inflatons). The process $\nu+\ol{\nu}\to \ell^- + \ell^+$  could result
in a comparable energy being ejected in highly relativistic, charged particles.
It is noteworthy that both the two-body decay to $\nu\ol{\nu}$, and
the above two-body reaction, can produce highly correlated 
(through coherence) streams of energy in opposite directions, in effect
a coherent entity moving outward from the source. Interactions
of the electrons with an existing, moderate atmosphere of ordinary matter
(say, of the order of $10^{13}$ particles/cm$^3$),
over a dimension considerably greater ($> 10^{10}\cm$) than that
of the very compact source, could produce the initial gamma-ray
burst within a short time interval $O({\mathrm{secs}})$. Subsequent interactions
would produce lower-energy photons in regions approaching the boundary
with space, over longer time intervals. The ``visibility''
(i.~e.~via gamma-rays and photons) of the burst, and the later
emission, ceases if the atmosphere is effectively dispersed by the 
high-energy collisions. However, the ``dark'' primary source is still producing
very high-energy neutrinos. There are clumps of dark matter which are not
associated with luminous centers or halos. Relative to the QSO's as discussed in this 
paper, the source mass is less by a factor of about $10^{-8}$
and the source dimension is also less by $\sim 10^{-8}$, being again
just above the Schwarzschild radius relevant to this mass. The
hypothetical decay of very massive dark matter provides a specific
connection between the energy sources of gamma-ray bursts and
of QSO's; this can produce a frequency for the former of 
$\simgt 10^{-6}$ per galaxy per year, and also some possible anisotropy
in the bursts at cosmological distances. The anisotropy would be mainly
evident in the GRB's which are nearer to us; this should be correlated
with these bursts being at higher energy, on the average\cite{ref11}.
This is natural consequence of the greater accretion of
the decay neutrinos around older systems, and of their more effective
conversion into the relativistic electrons which are essential for
bringing about a GRB.\footnote{For the converse reason, an
exception could be a lower-energy GRB which is correlated with
an unusually massive supernova event\cite{ref12}.}
\section*{Acknowledgement}
S.~B.~thanks Patrick Heiliger for much help.

\end{document}